\documentclass[12pt,a4paper]{article}
\pdfoutput=1
\usepackage{graphicx}
\usepackage[dvipsnames]{xcolor}

\usepackage{tikz}
\usetikzlibrary{decorations.pathmorphing}
\usetikzlibrary{decorations.pathreplacing}
\usetikzlibrary{arrows.meta}
\tikzset{
  >={To[length=5pt]}
  }
\usetikzlibrary{shapes, shapes.geometric, shapes.symbols, shapes.arrows, shapes.multipart, shapes.callouts, shapes.misc}
\tikzset{snake it/.style={decorate, decoration=snake}}
\tikzset{7brane/.style={circle, draw=black, fill=black,ultra thick,inner sep=1.5 pt, minimum size=1 pt,}, c/.default={4pt}}
\tikzset{cross/.style={cross out, draw=black,thick, minimum size=2*(#1-\pgflinewidth), inner sep=0pt, outer sep=0pt}, cross/.default={5pt}}
\tikzset{big7brane/.style={circle, draw=black, fill=black,ultra thick,inner sep=2.5 pt, minimum size=1 pt,}, c/.default={4pt}}
\tikzset{u/.style={circle, draw=black, fill=white,inner sep=2 pt, minimum size=2 pt,},f/.style={square, draw=black, fill=white,ultra thick,inner sep=4 pt, minimum size=2 pt,}}
\tikzset{so/.style={circle, draw=black, fill=red,inner sep=2 pt, minimum size=2 pt,},f/.style={square, draw=black, fill=white,ultra thick,inner sep=4 pt, minimum size=2 pt,}}
\tikzset{sp/.style={circle, draw=black, fill=blue,inner sep=2 pt, minimum size=2 pt,},f/.style={square, draw=black, fill=white,ultra thick,inner sep=4 pt, minimum size=2 pt,}}
\tikzset{uf/.style={rectangle, draw=black, fill=white,inner sep=3 pt, minimum size=4 pt,}}
\tikzset{spf/.style={rectangle, draw=black, fill=blue, thick,inner sep=3 pt, minimum size=4 pt, circle, draw=black, fill=blue,thick,inner sep=2 pt, minimum size=2 pt,},f/.style={square, draw=black, fill=white,ultra thick,inner sep=4 pt, minimum size=2 pt,}}
\tikzset{sof/.style={rectangle, draw=black, fill=red, thick,inner sep=3 pt, minimum size=4 pt,}}
\usetikzlibrary{positioning}
\usetikzlibrary{arrows}

\usepackage{jheppub}
\usepackage{amsmath,amssymb,euscript,array,mathrsfs,appendix,ctable,mathabx}
\usepackage{bm}
\usepackage[small]{caption}
\usepackage{xcolor}
\usepackage{float}
\usepackage{braket}
\usepackage{comment}
\usepackage{xcolor}
\usepackage{empheq}
\definecolor{lightblue}{RGB}{100,180,255}
\newcommand \arXiv [1]{\href{http://arxiv.org/abs/#1}{\tt arXiv:#1}}

\title{Holographic entanglement entropy and c-functions in conformal and confining backgrounds}

\author{Jonathan Whittle}

\affiliation{Centre for Quantum Fields and Gravity, \\
Department of Physics,\\
Swansea University,\\
Singleton Park, Swansea, SA2 8PP, United Kingdom}

\emailAdd{jonathan.whittle@swansea.ac.uk}

\abstract{In this work we investigate holographic spacelike and timelike entanglement entropy using the Ryu-Takayanagi prescription, for slab-shaped and ball-shaped entangling regions. We work with an infinite family of 10-dimensional Type IIB supergravity solutions, which are gravity duals to an infinite set of linear quiver theories, with the backgrounds defined using the electrostatic potential formalism for brane configurations. The dual theories are 4-dimensional confining theories at low energy, but decompactify and flow to 5-dimensional SCFTs in the UV. We find that the entanglement entropy exhibits phase transition behaviour, and we use our results to investigate proposed c-functions constructed from the entanglement entropy. Comparing with the flow central charge, another proposed c-function, we find that each displays good behaviour, and reflects both UV and IR features of the dual theory.}

\begin{document}
\maketitle
\flushbottom

\section{Introduction}

    The AdS/CFT correspondence initially supplied a duality between Type IIB string theory on AdS\textsubscript{5} times an $S^5$ and $\mathcal{N}=4$ SYM theory \cite{Maldacena:1997re,Gubser:1998bc,Witten:1998qj}. Since its conception many variations have been obtained \cite{Witten:1998zw,Polchinski:2000uf,Klebanov:2000hb,Maldacena:2000yy} with different backgrounds dual to different field theories (eg. confining theories and theories with varying amounts of SUSY preservation). One method for generating new backgrounds is to start from a lower-dimensional supergravity theory, find a solution, and then perform an uplift to some higher-dimensional string theory -- for example in this paper we look at a 10-dimensional Type IIB supergravity background which was obtained via uplift from 6-dimensional gauged Romans' supergravity \cite{Romans:1986er}. The field theory which is dual to this background is an infinite family of 5-dimensional quiver theories which are SCFTs in the UV but flow to gapped 4-dimensional non-supersymmetric theories in the IR. Each quiver theory is related to a specific brane configuration along one of the internal directions of the background -- here we employ the electrostatic potential formalism \cite{Legramandi:2021uds} which links the brane picture to its associated quiver theory. \\

    Entanglement entropy (EE) in a quantum system provides a measure of how entangled two separate regions are, and there are various methods in field theory to calculate it, however these can become difficult to work with in higher dimensions, and when dealing with theories which are not conformal. In these cases it is advantageous to employ holography, and compute the holographic entanglement entropy, pioneered in \cite{Ryu:2006bv}, which arises as a geometric quantity associated with the background dual to the field theory of interest. There is ongoing interest in holographic entanglement entropy for its potential as a tool to probe confinement, phase transitions, and RG flows \cite{Klebanov:2007ws,Kol:2014nqa,Jain:2020rbb}. Spacelike entanglement entropy, where the associated entangling region in the dual QFT is taken at a constant time slice, has been widely studied \cite{Rangamani:2016dms}, but currently timelike entanglement entropy \cite{Doi:2023zaf,Doi:2022iyj} and its holographic extension remain less explored. Some recent studies include \cite{Guo:2025pru,Afrasiar:2024lsi,Nunez:2025gxq,Jena:2024tly}. Typically when taking timelike entangling surfaces one ends up with a complex-valued entanglement entropy which can be interpreted as a pseudo-entropy \cite{Doi:2022iyj,Nakata:2020luh}, related to the entanglement associated with transitions between states. This also motivates an interpretation in terms of a complexified geometry \cite{Heller:2024whi,Heller:2025kvp} in which the embedding surface extends along a radial coordinate with both real and imaginary parts.\\
 
    The main goal of this paper is to compute and analyse holographic entanglement entropy for slab and spherical regions in a Type IIB background corresponding to a compactification of a 5-dimensional SCFT to a 4-dimensional gapped QFT. The background possesses both conformal (in the UV) and confining (in the IR) phases, the transition between which we investigate using the entanglement entropy. The analysis provides an explicit link between brane setups, quiver field theories, and geometric observables. We consider entanglement entropy for both spacelike and timelike entangling regions, and investigate phase transitions and c-functions. The presence of phase transitions is signposted by characteristic swallowtail behaviour in the EE, which can be a hallmark of confinement. We establish and compare two distinct c-function constructions, one from the EE and one from the 'flow central charge' which are monotonic and consistent with the physics of the dual QFT.\\

    In Section \ref{sec: background} we present the background and explain the electrostatic formalism for the brane setup and associated quiver theories. We expand on the dual field theory, and describe its key features in the UV and IR. In Section \ref{sec: review} we briefly review the concept of entanglement entropy and its holographic realisation, and then in Section \ref{sec: slab EE} we go on to compute the EE for slab-shaped entangling regions. We consider two different c-functions and check that they are physically consistent. In Section \ref{sec: sphere EE} we investigate the EE for spacelike spherical regions, and consider a candidate c-function. We comment on embeddings for timelike spherical regions. We conclude in Section \ref{sec: conclusion}.

\section{Background}\label{sec: background}

    We start with the Type IIB supergravity background constructed in \cite{Fatemiabhari:2024aua}, which belongs to a similar class of solutions as those studied in \cite{Anabalon:2021tua} (see \cite{Anabalon:2024che,Chatzis:2025dnu,Chatzis:2024kdu,Nunez:2023nnl,Nunez:2023xgl,Kumar:2024pcz,Barbosa:2024smw,Macpherson:2024qfi,Macpherson:2025pqi,Castellani:2024pmx,Castellani:2024ial} for related works on this type of background). It it parametrised by the coordinates $(t,x_1,x_2,x_3,r,\phi,\theta,\varphi,\sigma,\eta)$ and depends on the three parameters $(\tilde{g},\mu,c)$, which for now we will treat as arbitrary constants. Setting $\alpha'=g_s=1$ the metric is given by:

    \begin{align}\label{eqn: metric}
        ds^2 = & f_1 \Big[ \frac{2 \tilde{g}^2}{9} H^{1/2}(r) r^2 dx_{1,3}^2 + \frac{2 \tilde{g}^2}{9} \frac{H^{1/2}(r)}{f(r)} dr^2 + \frac{2 \tilde{g}^2}{9} H^{-3/2}(r) f(r) d\phi^2 \nonumber \\ & + f_2 \Big( d\theta^2 + \sin^2 \theta (d\varphi - A_1^{(3)})^2 \Big) + f_3 ( d\sigma^2 + d\eta^2) \Big]
    \end{align}

    where the two warp factors $f(r)$ and $H(r)$, and the gauge field component $A_1^{(3)}$ are given by

    \begin{equation}
        f(r) = -\frac{\mu}{r^3} + \frac{2 \tilde{g}^2}{9} r^2 H^2(r) \quad\quad\quad H(r) = 1 - \frac{c^2}{r^3} \quad\quad\quad A_1^{(3)} = \frac{\sqrt{2\mu}}{c} (1 - \frac{1}{H(r)}) d\phi.
    \end{equation}

    The spacetime asymptotically ($r \rightarrow \infty$) becomes AdS\textsubscript{6} times an $S^2$ fibred with the gauge field $A_1^{(3)}$, times a 2d Riemann surface parametrised by $(\sigma,\eta)$. This gives us an $SO(2,5) \times SO(3)$ isometry group. $\sigma$ ranges over $(-\infty,\infty)$, and $\eta$ is compact, taking values in the interval $[0,P]$.

    The presence of the warp factors means that $\phi$ parametrises an $S^1$ which decompactifies asymptotically but shrinks in the IR, and eventually reaches zero radius at a nonzero value of $r$. At this point, $r=r_*$ the space smoothly ends, which creates a cigar shape in $(r,\phi)$ with the topology of a disk. $r_*$ can be found from the largest real solution of $f(r_*)=0$. We also require $\phi$ to have the following periodicity to ensure the absence of conical singularities at $r = r_*$:

    \begin{equation}
        \phi \sim \phi + L_\phi \quad\quad \mathrm{with} \quad\quad L_\phi = 4\pi \frac{H(r_*)}{f'(r_*)}.
    \end{equation}

     We believe this corrects an error in equation 2.5 of \cite{Fatemiabhari:2024aua}. To arrive at this result, we expand the warp factors around $r_*$ to leading order in $(r-r_*)$, and then perform a coordinate transformation on $r$ to put the $(r,\phi)$ part of the metric into polar coordinate form ($dr^2 + \mathrm{const} \ r^2d\phi^2$). Requiring that we get $2\pi$ when integrating $\phi$ over its range $[0,L_\phi)$ leads us to the above result for $L_\phi$. \\

    The $f_i(r,\sigma,\eta)$ functions were first used in \cite{Legramandi:2021aqv}, for constructing uplifts to Type IIB of solutions to Romans' supergravity \cite{Romans:1986er}. Writing down explicitly all the fields appearing in the solution requires 7 different functions, but for us the only relevant ones are:

    \begin{equation}
        f_1 = \frac{3 \pi}{2 X^2} \Big( \frac{\sigma^2 \partial^2_\eta V + 3 X^4 \sigma \partial_\sigma V}{\partial^2_\eta V} \Big)^{1/2} \quad\quad f_2 = \frac{X^2 \partial_\sigma V \partial^2_\eta V}{3 \Lambda} \quad\quad f_3 = \frac{X^2 \partial^2_\eta V}{3 \sigma \partial_\sigma V}.
    \end{equation}

    We will also make use of the dilaton field $\Phi$, given by:

    \begin{equation}\label{eqn: dilaton}
        e^{-2 \Phi} = f_6 \quad\quad \mathrm{where} \quad\quad f_6 = \frac{36 X^4 \sigma^2 \partial_\sigma V \partial^2_\eta V}{(3X^4 \partial_\sigma V + \sigma \partial^2_\eta V)^2} \Lambda.
    \end{equation}

    The functions $\Lambda=\Lambda(\sigma,\eta)$ and $X=X(r)$ also have specific forms, but since they play no part in this analysis, we omit them here (the full functions, as well as the other $f_i$s, can be found in \cite{Fatemiabhari:2024aua}). \\
    
    The 'potential' $V(\sigma,\eta)$ is required to satisfy the following Laplace-like partial differential equation in order to satisfy the Type IIB equations of motion:

    \begin{equation}
        \partial_\sigma(\sigma^2 \partial_\sigma V) + \sigma^2 \partial_\eta^2 V = 0.
    \end{equation}

    Each choice of potential function satisfying the above requirement defines a member of an infinite family of asymptotically AdS\textsubscript{6} backgrounds. It is convenient to switch to $\hat{V}(\sigma,\eta) = \sigma V(\sigma,\eta)$, in terms of which the differential equation becomes $\partial_\sigma^2 \hat{V}+\partial_\eta^2\hat{V}=0$. The boundary conditions we impose are:

    \begin{align}
        & \hat{V}(\sigma \rightarrow \pm \infty, \eta)=0 , \quad \hat{V}(\sigma,\eta=0) = \hat{V}(\sigma,\eta=P)=0, \\ & \lim_{\epsilon \rightarrow0} \Big( \partial_\sigma \hat{V}(\sigma= +\epsilon,\eta) - \partial_\sigma \hat{V}(\sigma=-\epsilon,\eta) \Big) = \mathcal{R}(\eta)
    \end{align}

    where $\mathcal{R}(\eta)$ is a charge density called the rank function. These are the boundary conditions appropriate for the electrostatic potential between two parallel conducting plates positioned at $\eta=0$ and $\eta=P$ and extending in the $\sigma$ direction, with a charge distribution $\mathcal{R}(\eta)$ at $\sigma=0$ \cite{Legramandi:2021uds}. This is the reason that $V(\sigma,\eta)$ is referred to as a potential. \\

    The differential equation admits a solution which can be written as the following Fourier expansion, as shown in \cite{Legramandi:2021uds}:

    \begin{equation}\label{eqn: potential Fourier}
        \hat{V}(\sigma,\eta) = \sum^\infty_{k=1} a_k \sin \left( \frac{k \pi}{P}\eta \right) e^{-\frac{k\pi}{P}|\sigma|} \quad\quad \mathrm{where} \quad\quad a_k = \frac{1}{\pi k} \int^P_0 \mathcal{R}(\eta) \sin \left( \frac{k \pi}{P}\eta \right) \ d\eta
    \end{equation}

    so if we know the rank function, we can use it to reconstruct the potential. Quantisation of Page charges (see below) requires that the rank function is a convex polygonal of the form:

    \begin{equation}\label{eqn: rank function}
        \mathcal{R}(\eta) = \begin{cases} N_1 \eta \quad\quad & 0 \leq \eta \leq 1 \\ 
        N_l + (N_{l+1}-N_l)(\eta-l) \quad\quad & l \leq \eta \leq l+1 \quad \mathrm{for} \quad l=1, ..., P-2 \\ N_{P-1} (P-\eta) & (P-1) \leq \eta \leq P\end{cases}
    \end{equation}

    An example rank function is shown in Figure \ref{fig: rank function}. \\

    Page charges are calculated by integrating various fields appearing in the supergravity background over different submanifolds, and they are related to the number of branes present in the background. For example, NS5-branes couple to the 3-form field strength $H_3=dB_2$, and their corresponding Page charges are computed by integrating $H_3$ over some cycle in the geometry. The integral is evaluated using the explicit form of $B_2$ from the supergravity solution, and the result is:

    \begin{equation}
        Q_{NS5} = \frac{1}{4\pi^2} \int_{M_3} H_3 = P
    \end{equation}

    which counts the number of NS5-branes. Here $M_3$ is defined as $(\eta,S^2)$ with $\sigma \rightarrow \pm\infty$, $r \rightarrow \infty$.

    There are also Page charges associated with D7 and D5 branes:

    \begin{align}\label{eqn: brane charges}
        & Q_{D7}[k-1,k]=\mathcal{R}''(\eta)=2N_{k}-N_{k+1}-N_{k-1}\\ & Q_{D5}[k-1,k]=\mathcal{R}(\eta)-\mathcal{R}'(\eta)(\eta-k) = N_k
    \end{align}

    where $[k-1,k]$ refers to the interval in the $\eta$ coordinate. This means the rank function is related to the positions of various different stacks of branes in the $\eta$ direction. We can view this as the Hanany-Witten configuration \cite{Hanany:1996ie} displayed in Figure \ref{fig: Hanany Witten}. Ensuring that the Page charges are quantised (which they must be in order to have an integer number of branes) implies that the rank function is a convex, piecewise linear function with integer gradient for each interval \cite{Legramandi:2021uds}.
    
    Placing stacks of branes in the background is analogous to working with the rank function detailed above. The rank function in turn generates the potential by equation \ref{eqn: potential Fourier}, and then all of the $f_i$ functions on which the supergravity solution depends descend from this. On the field theory side, these stacks of branes relate to a quiver theory, which we now explain.

    \subsection{Dual field theory}\label{sec: dual FT}

    The stacks of D5-branes in the background house SQFTs with gauge group $SU(N_k)$ on their worldvolume, which are effectively (4+1)-dimensional since the branes span a finite interval in the $\eta$ direction, but are extended in the other 5 directions. The stacks of D7-branes are pointlike in the $\eta$ direction, and provide $SU(F_k)$ flavour groups for the D5-brane worldvolume theories. In brane models like this, usually the number of D5-branes is taken to be much larger than the number of D7-branes, which are then added in a probe approximation (also referred to as a 'quenched approximation'), however in this case the full back-reaction is accounted for (see \cite{Nunez:2010sf} for a review of brane backreaction). \\

    Taking the whole brane configuration into consideration, the resulting world-volume theory on the stacks of D5-branes is a linear, balanced quiver theory in 4+1 dimensions. Linear means that the gauge nodes are arranged in a single line which doesn't loop back on itself, and balanced means that the gauge and flavour nodes are related to each other by the balancing condition:

    \begin{equation}
        F_k=2N_k-N_{k-1}-N_{k+1}
    \end{equation}

    with gauge and flavour nodes $N_k$ and $F_k$ as shown in Figure \ref{fig: Hanany Witten}. This matches the expression for the D7-brane Page charges in terms of the D5-brane Page charges, equation \ref{eqn: brane charges}.

    In practice this means we can choose from an infinite family of quiver theories (up to the balancing condition) and then translate to the associated Hanany-Witten brane configurations on the gravity side. Each comes with a rank function, as in equation \ref{eqn: rank function}, from which we can obtain the potential $V(\sigma,\eta)$ via its Fourier expansion, equation \ref{eqn: potential Fourier}, and then all the $f_i$ functions present in the background follow. \\

    At high energies these quiver theories flow to a (4+1)-dimensional SCFT fixed point (for this to happen we actually only require the quiver to be underbalanced \cite{Intriligator:1997pq} in 5d, $F_k \leq 2N_k - N_{k+1}-N_{k-1}$, however if we want to make contact with the rank function coming from the brane setup, the stronger balancing condition must be satisfied). The SCFT can then be deformed by turning on relevant operators, which compactify the $\phi$ direction, and induce a flow down to a (3+1)-dimensional, non-supersymmetric, gapped QFT.
    
    We can find expectation values of operators on the field theory side by performing a near-boundary expansion of the bulk fields on the gravity side, and in doing so we find that the parameters $\mu$ and $c$ appearing in our two warp factors $f(r)$ and $H(r)$ are related to the vevs of the above mentioned relevant operators.
    
    Turning on $\mu$ we have a solution resembling that of Anabal\'{o}n and Ross in \cite{Anabalon:2021tua}, so we might assume/hope that some supersymmetry is preserved in the flow down to the 4d theory. Checking the SUSY transformations however reveals that unless $\mu=c=0$, supersymmetry is completely broken \cite{Fatemiabhari:2024aua}.

    \begin{figure}[t]
        \centering
        \begin{tikzpicture}[scale=1, every node/.style={scale=0.95}]
            
            \draw[black,line width=0.3mm] (1,0) -- (2,1.2);
            \draw[black,line width=0.3mm] (2,1.2) -- (3,2);
            \draw[black,line width=0.3mm] (3,2) -- (4,2.4);
            \draw[black,line width=0.3mm] (4,2.4)-- (5,2.2);
            \draw[black,line width=0.3mm]  (5,2.2)-- (6,1.4);
            \draw[black,line width=0.3mm] (6,1.4)-- (7,0);

            \draw (1,0) node[below] {$0$};
            \draw[gray,dashed,line width=0.3mm] (2,0)-- (2,1.2);
            \draw (2,0) node[below] {$1$};
            \draw[gray,dashed,line width=0.3mm] (3,0)-- (3,2);
            \draw (3,0) node[below] {$2$};
            \draw[gray,dashed,line width=0.3mm] (4,0)--  (4,2.4);
            \draw (4.3,-0.15) node[below] {$.~.~.$};
            \draw[gray,dashed,line width=0.3mm] (5,0)-- (5,2.2);
            \draw (6,0) node[below] {$P-1$};
            \draw[gray,dashed,line width=0.3mm] (6,0)--  (6,1.4);
            \draw (7,0) node[below] {$P$};
            
            \draw[gray,dashed,line width=0.3mm] (1,1.2)--  (2,1.2);
            \draw (1,1.2) node[left] {$N_1$};
            
            \draw[gray,dashed,line width=0.3mm] (1,2)-- (3,2);
            \draw (1,2) node[left] {$N_2$};
    
            \draw[-stealth, line width=0.53mm] (1,0)--(1,3) node[left ]{$\mathcal{R}(\eta)$};
            \draw[-stealth, line width=0.53mm] (1,0) --(7.6,0) node[below ]{$\eta$};
        \end{tikzpicture}
        \caption{An example rank function, which is continuous, convex, and piecewise linear as required by the quantisation of Page charges. The ranks of the associated gauge groups are encoded in the values of $\mathcal{R}(\eta)$ at the points where the gradient is discontinuous. Note that the rank function does not have to be symmetric.}
        \label{fig: rank function}
    \end{figure}

    \begin{figure}[t]
        \centering
        \begin{scriptsize}
        \begin{tikzpicture}[scale=.8]
        \node at (4,-2){(a)};
       
        \draw[thick](0,-.5)--(0,2.5);
        \node[label=below:{NS5$_1$}]at(0,-.5){};
       \node[label=above:{$F_2$ D7}][cross] at(3,1.8) {};
       \node[label=above:{$F_{p-1}$ D7}][cross] at(7,1.8) {};
       \node[label=above:{$F_1$ D7}][cross] at(1,1.8) {};
       \draw[thick](2,-.5)--(2,2.5);
       \node[label=below:{NS5$_2$}]at(2,-.5){};
       \node[label=below:{NS5$_3$}]at(4,-.5){};
       \node[label=below:{NS5$_{P-1}$}]at(6,-.5){};
       \node[label=below:{NS5$_P$}]at(8,-.5){};
       \node at (5,1.25) {$\cdots$};
       \draw[thick](4,-.5)--(4,2.5);
       \draw[thick](8,-.5)--(8,2.5);
       \draw[thick](6,-.5)--(6,2.5);
       \draw[thick](0,1)--(4,1);
       \node at (1,.6) {$N_1$ D5};
        \draw[thick](6,1)--(8,1);
        \node at (3,.6) {$N_2$ D5};
        \node at (7,.6) {$N_{p-1}$ D5};
       \end{tikzpicture}
       \end{scriptsize}
       \hspace{1cm}
       \begin{scriptsize}
                 \begin{tikzpicture}
                    \node at (-1,-2){(b)};
                    
                          \node (1) at (-3,1.1) [circle,draw,thick,minimum size=1cm] {N$_1$};
                        \node (2) at (-1.5,1.1) [circle,draw,thick,minimum size=1cm] {N$_2$};
                        \node (3) at (0,1.1)  {$\dots$};
                        \node (5) at (1.5,1.1) [circle,draw,thick,minimum size=1cm] {N$_{P-1}$};
                        \draw[thick] (1) -- (2) -- (3) -- (5);
                        \node (1b) at (-3,-0.9) [rectangle,draw,thick,minimum size=0.8cm] {F$_1$};
                        \node (2b) at (-1.5,-0.9) [rectangle,draw,thick,minimum size=0.8cm] {F$_2$};
                        \node (3b) at (0,-0.9)  {$\dots$};
                        \node (4b) at (1.5,-0.9) [rectangle,draw,thick,minimum size=0.8cm] {F$_{P-1}$};
                        \draw[thick] (1) -- (1b);
                        \draw[thick] (2) -- (2b);
                        \draw[thick] (5) -- (4b);
                 \end{tikzpicture}
             \end{scriptsize}
      
        \caption{The Hanany-Witten brane setup is shown in (a), with the horizontal direction corresponding to $\eta$. Each stack of D5-branes is suspended between two NS5-branes, and there are transverse stacks of D7-branes providing flavour groups. In (b) is the corresponding quiver plot. Gauge nodes come from the stacks of D5-branes, and flavour nodes come from the stacks of D7-branes.}
        \label{fig: Hanany Witten}
    \end{figure}

\section{Holographic entanglement entropy review}\label{sec: review}

    We now give a brief introduction to entanglement entropy and the method for its computation via holography.

    For a quantum mechanical system in a state $|\psi\rangle$ the density matrix is given by $\rho=|\psi\rangle \langle \psi |$ and the Von Neumann entropy is then defined by:

    \begin{equation}
        S= - \mathrm{tr} ( \rho \ln \rho).
    \end{equation}

    If we divide the system into two regions, $A$ and $B$, we can find the reduced density matrix for region $A$, $\rho_A= \mathrm{tr}_{B} (\rho)$ by tracing over region $B$. More specifically we divide the Hilbert space up as the direct product $\mathcal{H} = \mathcal{H}_A \otimes \mathcal{H}_B$ and then take the trace over $\mathcal{H}_B$. The entanglement entropy is then found from the Von Neumann entropy for one of the regions (if we have a region and its complement, then $S_{EE,A}=S_{EE,A'}$), so for region $A$:

    \begin{equation}
        S_{EE} = - \mathrm{tr}(\rho_A \ln \rho_A).
    \end{equation}

     The entanglement entropy provides a measure of how entangled the two regions are. It also obeys various subadditivity properties which we will mention in Section \ref{sec: c function}. The standard way to compute EE in field theory is via the replica trick -- this involves evaluating $\mathrm{tr}_A \rho_A^n$, then differentiating with respect to $n$, and then sending $n \rightarrow 1$ (see \cite{Rangamani:2016dms} for a review).\\

     We should also mention the concept of pseudo-entropy, which will become relevant when we discuss timelike EE. If we have two pure states $| \psi \rangle$ and $|\varphi \rangle$ then we can define the reduced transition matrix $\tau_A$ by:

     \begin{equation}
        \tau_A = \mathrm{tr}_B \left( \frac{| \psi \rangle \langle \varphi|}{\langle \varphi | \psi \rangle} \right).
     \end{equation}

     The pseudo-entropy for region $A$ is then defined as the Von Neumann entropy associated with the transition matrix:

     \begin{equation}
        S_P = - \mathrm{tr}( \tau_A \ln \tau_A )
     \end{equation}

     which corresponds to the entanglement associated with transitions between states. This quantity is in general complex-valued.\\

    The entanglement entropy becomes difficult to compute in field theory in dimensions greater than 2. Motivated by the Bekenstein-Hawking entropy, which is calculated from the area of a black hole horizon, and the analogous dividing up of space into two completely separate regions (the inside and outside of the black hole horizon), Ryu and Takayanagi (RT) proposed a holographic method for calculating entanglement entropy \cite{Ryu:2006bv, Ryu:2006ef}. Their proposal is that for a CFT in $d$ dimensions, the entanglement entropy for a region $A$ is computed as a geometric quantity using the following area law:

    \begin{equation}
        S_{EE} = \frac{\mathrm{area}(\gamma_A)}{4 G^{d+1}_N}
    \end{equation}

    where $\gamma_A$ is a $(d-1)$-dimensional minimal surface living in AdS\textsubscript{$d+1$}, attached to $\partial A$ on the AdS boundary, and $G_N^{d+1}$ is the $(d+1)$-dimensional Newton's constant. As we will find, there can be more than one minimal surface for a given choice of $\partial A$. We now use the RT prescription to calculate entanglement entropies for different-shaped regions in the dual field theories described above.
    
\section{Entanglement entropy for slab regions}\label{sec: slab EE}

    In this section we will calculate the entanglement entropy of a slab region in the dual field theory -- this means a region which is finite in some directions (in this case $t$ and $x_1$) and extended in all the others. We use the Ryu-Takayanagi prescription to obtain the EE via calculation of the area of an embedding surface which we now describe. \\

    The embedding is obtained by locating the slab entangling region $A$ on the AdS boundary $r \rightarrow \infty$. We then attach an 8-surface to the boundary $\partial A$ of the entangling region, and allow it to delve into the bulk. We choose the 8-surface $\Sigma$ parametrised by ($x_2,x_3,r,\phi,\theta,\varphi,\sigma,\eta$) and take $t=t(r)$ and $x_1=x_1(r)$. The induced metric on the 8-surface is:

    \begin{align}
        ds_{\mathrm{ind}}^2 = & f_1 \Big[ \frac{2 \tilde{g}^2}{9} H^{1/2}(r) ( -t'^2 + x_1'^2 + \frac{1}{r^2 f(r)}) r^2 dr^2 + \frac{2 \tilde{g}^2}{9} H^{1/2}(r) r^2 (dx_2^2 + dx_3^2) \nonumber \\ & + \frac{2 \tilde{g}^2}{9} H^{-3/2}(r) f(r) d\phi^2 + f_2 (d\theta^2 + \sin^2\theta d\varphi^2) + f_3 (d\sigma^2 + d\eta^2) \Big]
    \end{align}

    where a prime indicates differentiation with respect to $r$.  The entanglement entropy can be computed holographically according to Ryu-Takayanagi by minimising:

    \begin{equation}\label{eqn: EE formula}
        S_{EE} = \frac{1}{4 G_{10}} \int_\Sigma \sqrt{e^{-4\Phi} \det(g_{\mathrm{ind}})}
    \end{equation}

    where $G_{10}$ is the 10-dimensional Newton's constant, and we are working in string frame, so the factor of $e^{-4\Phi}$ ensures that the quantity inside the square root is a U-duality invariant. Expanding the determinant and substituting the dilaton solution from \ref{eqn: dilaton} we find:

    \begin{equation}
        e^{-4\Phi} \det(g_{\mathrm{ind}}) = f_1^8 f_2^2 f_3^2 f_6^2 \Big( \frac{2 \tilde{g}^2}{9} \Big)^4 r^6 f(r) (- t'^2 + x_1'^2 + \frac{1}{r^2 f(r)}) \sin^2 \theta.
    \end{equation}

    Expanding all the $f_i$s, and plugging this into equation \ref{eqn: EE formula} we obtain:

    \begin{equation}\label{eqn: square root eqn}
        4 G_{10} S_{EE} = 3^4 \pi^5 \Big( \frac{2 \tilde{g}^2}{9} \Big)^2 L_2 L_3 L_\phi \int d\sigma \int d\eta \ \sigma^3 \partial_\sigma V \partial_\eta^2 V \int dr \sqrt{r^6 f(r) (-  t'^2 + x_1'^2) + r^4}
    \end{equation}

    where $L_2$ and $L_3$ are the lengths of the $x_2$ and $x_3$ directions respectively. We'll split this up so that everything not included in the integral over $r$ on the RHS we denote by $\hat{\mathcal{N}}$:

    \begin{equation}
        \hat{\mathcal{N}} = 3^4 \pi^5 \Big( \frac{2 \tilde{g}^2}{9} \Big)^2 L_2 L_3 L_\phi \int d\sigma \int d\eta \ \sigma^3 \partial_\sigma V \partial_\eta^2 V.
    \end{equation}

    Evaluating $\hat{\mathcal{N}}$ requires knowledge of the specific quiver theory one is working with -- as described in Section \ref{sec: dual FT} the potential descends from the specific rank function we start with. Here we will not concern ourselves with this, being mainly interested in quantities' $r$-dependence, so we will usually write $S_{EE}/\hat{\mathcal{N}}$ on the left hand side of equations, and it is understood that 'entanglement entropy' refers to this quantity as well. The splitting up in this way of the expressions for the EE means that information about the specific quiver theory we're working with is contained only in $\hat{\mathcal{N}}$, and is subsequently isolated from our analysis, which is concerned with the $r$-dependent piece. This kind of universal behaviour is detailed in \cite{Chatzis:2025dnu}.\\

    Now we want to minimise the area of our embedding surface. For slab regions our embedding surface has a nontrivial profile in only one direction, so this calculation proceeds in a similar fashion to how one might calculate Wilson loops in holography -- the quantity inside the $r$ integral is our Lagrangian from which we derive equations of motion, and then the solutions $t(r)$, $x_1(r)$, define the minimal embeddings. A general method for obtaining the quark separation and quark-antiquark energy in Wilson loop calculations as functions of the embedding surface turning point was presented in \cite{Nunez:2009da}, which can be straightforwardly adapted to suit our purpose. We pick out the functions:

    \begin{equation}\label{eqn: GandF}
        F^2(r) = r^6 f(r) \quad\quad G^2(r) = r^4
    \end{equation}

    from the square root in equation \ref{eqn: square root eqn}, with which we can write the equations of motion for $t$ and $x_1$ as:

    \begin{equation}
        \frac{- F^2 t'}{\sqrt{F^2(- t'^2 + x'^2)+G^2}}= c_t \quad\quad \frac{F^2 x_1'}{\sqrt{F^2(- t'^2 + x'^2)+G^2}}= c_{x_1}
    \end{equation}

    where $c_t$ and $c_{x_1}$ are constants coming from the fact that the Lagrangian has no explicit $t$ or $x_1$ dependence. Rearranging, we find:

    \begin{equation}
        t'^2 = \frac{G^2 c_t^2}{F^2(F^2 - (- c_t^2 + c_{x_1}^2))} \quad\quad x_1'^2 = \frac{G^2 c_{x_1}^2}{F^2 (F^2 - (- c_t^2 + c_{x_1}^2)) }
    \end{equation}

    so the turning point $r_0$, at which $t'$ and $x_1'$ diverge, has $F^2(r_0) = - c_t^2 + c_{x_1}^2$. This means that the turning point depends on our choice of $(c_x,c_t)$ (or conversely, a given turning point restricts the allowed values of $c_x$ and $c_t$). We see that if we want to consider embeddings which depend only on $t$ or $x_1$ we can achieve this by setting $c_{x_1}=0$ or $c_t=0$ respectively. To obtain the separations we simply integrate the above, which gives us:
    
    \begin{equation}
        T = 2c_t \int^\infty_{r_0} dr \frac{G(r)}{F(r) \sqrt{F^2(r) - F^2(r_0)}} \quad\quad X_1 = 2c_{x_1} \int^\infty_{r_0} dr \frac{G(r)}{F(r) \sqrt{F^2(r) - F^2(r_0)}}
    \end{equation}
    
    where the factor of 2 is because we're only integrating over half of the surface. The squared interval $\Delta^2=X_1^2-T^2$ tells us whether the entangling region is spacelike ($\Delta^2>0$) or timelike ($\Delta^2 <0$). The regularised entanglement entropy can be expressed in terms of the $F$ and $G$ functions like:

    \begin{equation}
        \frac{4 G_{10} S_{EE}}{\hat{\mathcal{N}}} = 2 \int^\infty_{r_0} dr \frac{G(r) F(r)}{\sqrt{F^2(r) - F^2(r_0)}} - 2 \int^\infty_{r_*} dr \ G(r).
    \end{equation}

    where the second term is necessary for regularising UV divergences and corresponds to removing those embeddings where the surface hangs straight down from the AdS boundary to the end of space at $r=r_*$.

    Substituting in our expressions for $F$ and $G$ (equation \ref{eqn: GandF}), we end up with:

    \begin{equation}\label{eqn: separation}
        T = 2 c_t \int^\infty_{r_0} dr \frac{1}{r \sqrt{f(r)} \sqrt{r^6 f(r) - r_0^6 f(r_0)}}
    \end{equation}

    and the same expression for $X_1$, but with $c_t$ exchanged for $c_{x_1}$. The entanglement entropy is:

    \begin{equation}\label{eqn: EE}
        \frac{4 G_{10} S_{EE}}{\hat{\mathcal{N}}} = 2 \int^\infty_{r_0} dr \frac{r^5 \sqrt{f(r)}}{\sqrt{r^6 f(r) - r_0^6 f(r_0)}} - 2 \int^\infty_{r_*} dr \ r^2.
    \end{equation}

\begin{figure}
    	\centering
    	\includegraphics[width = 6.5in]{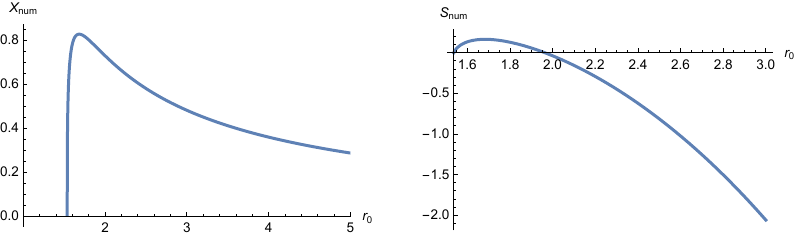}
    	\caption{The spacelike separation and entanglement entropy from equations \ref{eqn: separation} and \ref{eqn: EE}, with the integrals evaluated numerically. We vary $c_{x_1}$ and then plot parametrically $X_1(c_{x_1})$ and $S(c_{x_1})$ against $r_0(c_{x_1})$. We make the parameter choice $(\tilde{g},\mu,c) = (1,1,1)$ which puts the end of space at $r_* \approx 1.54$. The separation increases from 0 at $r_0=r_*$ to some maximum value, then asymptotes back to 0 meaning every value of $X$ has two different associated turning points.}
   	\label{fig: numerical separation and EE}
    \end{figure}

    These integrals cannot be performed analytically, but we can still make numerical plots of these quantities as functions of the turning point $r_0$, as in Figure \ref{fig: numerical separation and EE}. We find that multiple values of $r_0$ can lead to the same separations ($T$ or $X_1$), which could indicate a phase transition -- if we go on to plot the EE as a function of the separation (Figure \ref{fig: cusps}) we obtain a distinctive swallowtail-like cusp which is a strong indication of a phase transition. Although we might expect a deconfinement phase transition we cannot conclude from the behaviour of the entanglement entropy that this is what is being observed \cite{Jokela:2020wgs,Kol:2014nqa}. Instead one must perform the standard Wilson loop analysis and demonstrate an area law.

\subsection{Reality of the turning point}\label{sec: reality of TP}

    When calculating timelike entanglement entropy it is often the case that the embedding surface ends up with a complex-valued turning point \cite{Doi:2023zaf,Heller:2024whi}. Let's consider entangling regions which are purely timelike, so $c_{x_1}=0$. The turning point can be found by solving:

    \begin{equation}\label{eqn: turning point eqn}
        F^2(r_0) = -c_t^2 \quad \implies \quad \frac{2\tilde{g}^2}{9} r_0^8 - \frac{4\tilde{g}^2}{9} c r_0^5 - \mu r_0^3 + \frac{2\tilde{g}^2}{9} c^2 r_0^2 + c_t^2 = 0.
    \end{equation}

    For $c=\mu=0$ this equation has solutions only for complex turning points. This is the case if our background is pure AdS, however the addition of the warp factors changes this. In \cite{Nunez:2025ppd} it was found that backgrounds with a confining scale (for us this is supplied by $f(r)$) can have real turning points even for purely timelike entangling regions, and they obtained the turning points' explicit forms. In our case we should recover a similar result, however our turning point equation is more complicated due to the extra warp factor, so we will have to be content with searching numerically for the existence of real turning points. \\

    We should first observe that whenever we have a real-valued turning point it will necessarily be behind the end of space. Because $r_*$ is defined as the largest real root of $f(r)=0$ this means that $f(r)$ is positive for $r > r_*$, and tends to $\frac{2 \tilde{g}}{9} r^2$ as $r \rightarrow \infty$. This means in turn that $r_0^6 f(r_0) + c_t^2 = 0$ has real solutions only when $f(r_0) < 0$, or equivalently, when $r_0 < r_*$. This means that the embeddings will hang down from the boundary, reach the end of space, and lie flat along it for some distance -- this creates a 'bathtub' configuration. \\
    
    If we switch on $\mu$ while keeping $c=0$ we can obtain real turning points for $|\mu| > \mu_c$ where $\mu_c$ is some critical value. The equation is still quite hard to solve, but we can find $\mu_c$ by considering one of the minima, and checking at which $\mu$ value it moves below 0. We find that for $c=0$, the critical value is given by:

    \begin{equation}\label{eqn: critical vals}
        \mu_c = 8 \left( \frac{2\tilde{g}^2}{9} \right)^{3/8} \left( \frac{c_t^{10}}{3^3 5^5} \right)^{1/8}.
    \end{equation}

\begin{figure}
    	\centering
    	\includegraphics[width = 4in]{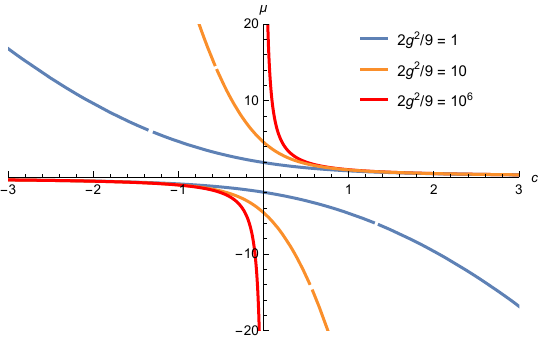}
    	\caption{Here we take three different values for $2\tilde{g}^2/9$ with $c_t=1$ and plot the values of $\mu$ and $c$ for which equation \ref{eqn: turning point eqn} has a single solution (the critical $\mu$ values mentioned above). The interior region (between each pair of lines) is the parameter space for which there are no real solutions to equation \ref{eqn: turning point eqn}. The shape of the plot matches our observations.}
   	\label{fig: aeroplane plot}
    \end{figure}

    If we turn $c$ on while keeping $\mu=0$, a second minimum appears but it can be shown to never go below 0 so we're stuck with complex solutions. We can then turn $\mu$ on and, for each value of $c$, there appear upper and lower critical $\mu$ values above and below which we obtain real solutions. These critical values are difficult to obtain analytically. In Figure \ref{fig: aeroplane plot} we plot numerically the critical values, and we can identify the interior region as the parameter space where no real turning point can be found.

    From equation \ref{eqn: critical vals} we can read off the intercepts of the $\mu$ axis, and we see that sending $\tilde{g}^2 \rightarrow \infty$ makes the critical value $\mu_c$ diverge. Plotting the critical values for multiple different choices of $2\tilde{g}^2/9$ (as in Figure \ref{fig: aeroplane plot}) we see that increasing this parameter causes the plot to behave like $\mu = c^{-(2n+1)}$ for $n \in \mathbb{Z}$.

\subsection{Approximate expressions}

    Based on the similarity with Wilson loop calculations, in \cite{Nunez:2025puk} approximate expressions which we can use for the separation and entanglement entropy were proposed. For the choice $c_t = 0$ they are given by:

    \begin{equation}
        X_{\mathrm{app}}(r_0) = \pi \frac{G(r_0)}{F'(r_0)} \quad \quad S_{EE \ \mathrm{app}}(r_0) = \pi \int F(r_0) \frac{d}{dr_0}\Big( \frac{G(r_0)}{F'(r_0)}\Big) dr_0 + \mathrm{const}.
    \end{equation}

\begin{figure}
    	\centering
    	\includegraphics[width = 6.5in]{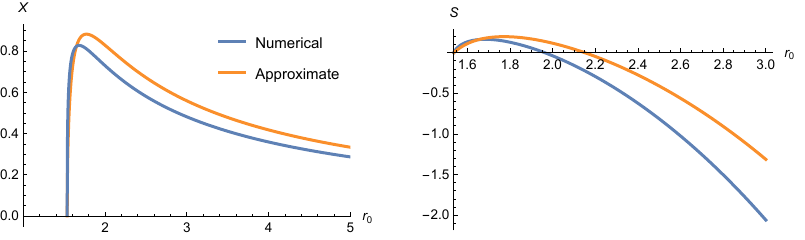}
    	\caption{The approximate separation and EE given in equation \ref{eqn: approximates} compared to the analytic expressions in equation \ref{eqn: separation} and equation \ref{eqn: EE} (numerically integrated) respectively. When comparing the EEs on the right we have shifted the approximate expression by a constant (which we expect based on \ref{eqn: approximates}) so that both are zero at $r_0=r_*$. Both panels show good qualitative agreement but they seem to differ by a multiplicative constant.}
   	\label{fig: separation and EE}
    \end{figure}

\begin{figure}
    	\centering
    	\includegraphics[width = 6.5in]{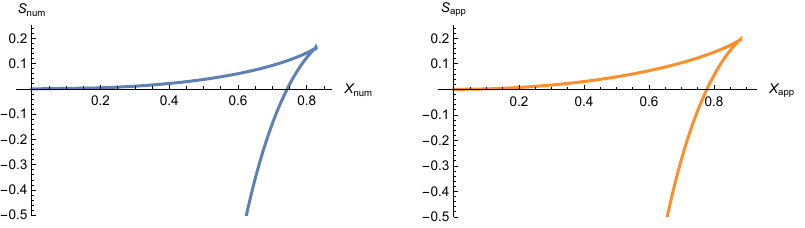}
    	\caption{The separation plotted against the EE. On the left, the analytic expressions with integrals evaluated numerically are used, and we find a cusp indicating a phase transition. On the right, the approximate expressions lead to a similar cusp showing good qualitative agreement.}
   	\label{fig: cusps}
    \end{figure}

    Because these take a much simpler form than our above results, we can evaluate them analytically. Using our expressions for $F(r)$ and $G(r)$ in equation \ref{eqn: GandF}, these come out as:

    \begin{equation}\label{eqn: approximates}
        X_{\mathrm{app}}(r_0) = \frac{2 \pi \sqrt{f_0}}{6f_0+r_0f'_0} \quad\quad S_{EE \ \mathrm{app}}(r_0) = \pi \int r_0^3 \frac{-8 f_0 f'_0 + r_0 f'^2_0 - 2 r_0 f_0 f''_0}{(6f_0 + r_0 f'_0)^2} \ dr_0 + \mathrm{const}
    \end{equation}

    where we have abbreviated $f(r_0)=f_0$. Plotting the approximate against the analytic expressions (numerically integrated) for the separation and entanglement entropy, in Figure \ref{fig: separation and EE}, we find good qualitative agreement. Plotting $X_\mathrm{app}$ against $S_{EE \ \mathrm{app}}$ we obtain the cusp in the right panel of Figure \ref{fig: cusps}. The two quantities have the same turning point by construction, which ensures a sharp cusp.

\subsection{C-function from entanglement entropy}\label{sec: c function}

    As mentioned in Section \ref{sec: review}, the entanglement entropy obeys various subadditivity properties. For example, if we have two spacelike entangling regions $A$ and $B$, then the entanglement entropy obeys the strong subadditivity inequality \cite{Lieb:1973cp}:

    \begin{equation}
        S_{EE}(A) + S_{EE}(B) \geq S_{EE}(A \cap B) + S_{EE}(A \cup B)
    \end{equation}

    which is saturated when the boundary of $B$ lies on the light cone of the boundary of $A$. \\

    The subadditivity properties mean that the EE is a quantity which lends itself to the construction of c-functions, measures of the degrees of freedom in a system. And indeed the equalities presented in \cite{Casini:2017vbe} imply that certain c-functions in $d=2$ and $d=3$ obey monotonicity -- they decrease monotonically from the UV to the IR. This motivates the search for c-functions constructed from the entanglement entropy of 4-dimensional theories \cite{Myers:2012ed,Jokela:2025cyz}, and in this section we will consider one example. \\

\begin{figure}
    	\centering
    	\includegraphics[width = 4in]{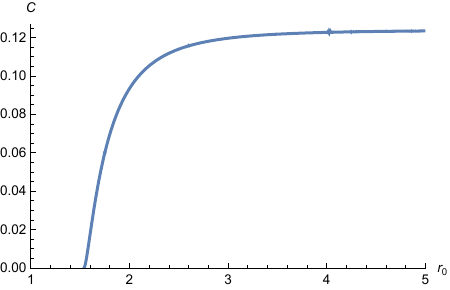}
    	\caption{The proposed c-function from equation \ref{eqn: c function}, plotted using the analytic expressions for $X_1$ and $S_{EE}$, numerically evaluated. It increases monotonically from the end of space $r_*$ (here at $r \approx 1.54$) as a good c-function should. The flattening out reflects the fact that the theory approaches a UV fixed point at high energy.}
   	\label{fig: slab c function}
    \end{figure}

    In \cite{Myers:2012ed} it was suggested that we can use the entanglement entropy of a slab region to obtain a c-function in the following way:

    \begin{equation}\label{eqn: c function}
        C(X, L) = \frac{X^{d-1}}{2^{d-2} L^{d-2}} \partial_X S(T,L) = \frac{X^4}{8 L^3}\partial_X S(X,L)
    \end{equation}

    where the width of the slab is $X$ (we are again taking $c_t =0$) and $L$ is the length of the other extended directions (so we take $L \rightarrow \infty$). This means that the slab is ($d-1$)-dimensional, so for us $d-1=4$. Plotting this using our expressions for the approximate separation and EE, we obtain Figure \ref{fig: slab c function}.

    We see that the value decreases monotonically from the UV to the IR, and becomes zero at $r=r_*$ where the space ends. This means that this c-function reflects the gapped nature of the dual theory, which is something we would hope for when constructing a c-function -- below some energy scale in the dual field theory (some radius on the gravity side) there are no degrees of freedom. It also seems to approach some maximum value as we increase the turning point away from the confining scale. This is what we would expect if the dual field theory approaches a UV fixed point where the central charge stays constant, and indeed as mentioned in Section \ref{sec: dual FT} our theory flows to a 5d SCFT fixed point at high energies. From these considerations we should conclude that the c-function in equation \ref{eqn: c function} is a good c-function. \\

    At this point we should recall that our background features a compact dimension $\phi$ in the bulk, which decompactifies as $r \rightarrow \infty$. This is a consequence of the presence of the warp factor $f(r)$, and it means that in the IR we have an effective 4-dimensional theory which becomes 5-dimensional in the UV. The flow between different dimensions introduces extra degrees of freedom into our system from the tower of KK modes associated with the compact dimension. As we increase the energy scale into the UV, more and more of the levels become accessible and this corresponds to a linear increase in the number of degrees of freedom of the system, as seen by a lower-dimensional observer. This doesn't mean that there are actually any 'new' degrees of freedom entering the system, it just reflects the fact that we have a compact dimension.

    We might expect that our c-function would reflect this, and end up tending to a linear function of $r$ in the UV to reflect the linear increase in degrees of freedom (visible to the lower-dimensional observer). It is interesting that the c-function doesn't display this behaviour and instead tends to a constant value. This suggests that this c-function somehow knows simultaneously about the gapped 4-dimensional IR theory, and also the 5-dimensional SCFT UV fixed point. It is interesting that the calculation of EE makes no distinction between internal and external coordinates, but the c-function seems to.

\subsection{Comparison with flow central charge}\label{sec: flow central charge}

    The central charge is a quantity which is only properly defined at conformal points, so when we move away from the AdS boundary, we flow away from the UV fixed point on the field theory side and the central charge can become harder to work with. The problem mentioned above is a key issue -- how to deal with the apparent new degrees of freedom introduced by the flow between dimensions. In \cite{Bea:2015fja} a way to compensate for this possibility was introduced. First, the metric must be put into the form:

    \begin{equation}
        ds^2 = -\alpha_0 dt^2 + \alpha_1 dy_1^2 + \alpha_2 dy_2^2 + ... + \alpha_d dy_d^2 + (\alpha_1 \alpha_2 ... \alpha_d)^{1/d} b(r) dr^2 + g_{ij}(d\theta^i - A^i_1)(d\theta^j - A^j_1)
    \end{equation}

    and then we consider a submanifold spanned by the $y$s and the $\theta^i$s. We form the combination:

    \begin{equation}
        V_\mathrm{sub} = \int_X \sqrt{e^{-4\Phi}\mathrm{det}[g_\mathrm{sub}]} \quad\quad \hat{H} = V_\mathrm{sub}^2
    \end{equation}

    where $g_\mathrm{sub}$ is the metric on the submanifold, and the integral is performed over the internal space $X$ spanned by the $\theta^i$s. The 'flow central charge' is then given by \cite{Merrikin:2022yho}:

    \begin{equation}\label{eqn: flow central charge}
        c_\mathrm{flow} = \frac{d^d}{G_{10}} b^{d/2}(r) \frac{\hat{H}^{\frac{2d+1}{2}}}{\hat{H}'^d}
    \end{equation}

    where the prime represents differentiation with respect to $r$. Picking out the $\alpha$s and $b(r)$ from our metric in equation \ref{eqn: metric} we substitute them into the above to find:

    \begin{equation}
        V_\mathrm{sub} = \hat{\mathcal{N}} r^3 \sqrt{f(r)} \quad\quad \hat{H}=\hat{\mathcal{N}}^2 r^6 f(r) \quad\quad \implies \quad\quad c_\mathrm{flow} = \left( \frac{2}{3} \right)^4 \frac{\hat{\mathcal{N}}}{G_{10}} \frac{H(r) r^4 f(r)^2}{(f(r)+ \frac{r f'(r)}{6})^4}.
    \end{equation}

\begin{figure}
    	\centering
    	\includegraphics[width = 4in]{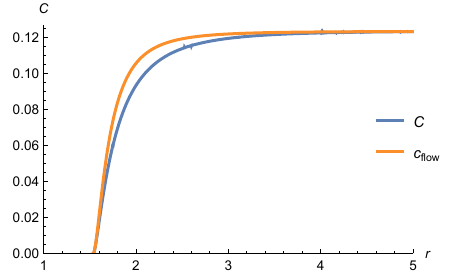}
    	\caption{The flow central charge from equation \ref{eqn: flow central charge} (in orange) compared with the c-function derived from the EE (in blue). Both increase monotonically from 0 at the end of space, up to some constant value, which is the expected behaviour for a theory gapped in the IR, and possessing a UV fixed point. The two c-functions differ in normalisation, which we have corrected by a rescaling of $c_{\mathrm{flow}}$ for ease of comparison.}
   	\label{fig: flow central charge}
    \end{figure}

    Plotting the expression for $c_\mathrm{flow}$ against $r$ we obtain Figure \ref{fig: flow central charge}. Once again the function is monotonic, and vanishes at the end of space $r=r_*$. It also plateaus, as before, revealing the UV fixed point once the degrees of freedom coming from the flow between dimensions have been accounted for. \\

    Obtaining the flow central charge requires a fairly unintuitive calculation, despite the fact that we end up with a nice and well-behaved c-function. For this reason c-functions derived from the entanglement entropy, like that in Section \ref{sec: c function}, are a much more desirable option -- the EE is an observable quantity allowing for the construction of a c-function in a covariant way.

\section{Entanglement entropy for spherical regions}\label{sec: sphere EE}

    Now we will consider the case when the entangling region is a ball on the AdS boundary, so that its boundary is a sphere. This sphere will then become the anchor to which we will attach our embedding surface. Spherical regions can exhibit more interesting properties than their slab shaped counterparts \cite{Jokela:2025cyz}, because as we increase the size of the region both its area and volume will increase (whereas for slabs only the volume increases).
    
    To better identify the ball region we'll be focusing on we can rewrite the metric in equation \ref{eqn: metric} as:

    \begin{align}\label{eqn: ball metric}
        ds^2 = & f_1 \Big[ \frac{2 \tilde{g}^2}{9} H^{1/2}(r) r^2 \Big( - dt^2 + d\rho^2 + \rho^2 d\Omega_2^2\Big) + \frac{2 \tilde{g}^2}{9} \frac{H^{1/2}(r)}{f(r)} dr^2 + \ ...
    \end{align}

    This time we'll choose the 8-surface $\Sigma$ to be parametrised by ($\Omega_2,r,\phi,\theta,\varphi,\sigma,\eta$) and we take $t=t(r)$ and $\rho = \rho(r)$. The U-duality invariant gains a factor of $\rho^4$ and we also pick up $4 \pi$ from the volume of the $\Omega_2$ 2-sphere when integrating. This means the entanglement entropy will be obtained this time by minimising:

    \begin{equation}
        4 G_{10} S_{EE} = 3^4 \pi^5 (4\pi) L_\phi \Big( \frac{2 \tilde{g}^2}{9} \Big)^2 \int d\sigma \int d\eta \ \sigma^3 \partial_\sigma V \partial_\eta^2 V \int dr \ r^2 \rho^2 \sqrt{f(r) r^2 (- t'^2 + \rho'^2) + 1}
    \end{equation}

    and again we'll relabel everything outside of the integral over $r$ as $\hat{\mathcal{N}}$, and then forget about it. The quantity inside the integral we take as a Lagrangian for $\rho(r)$, and we'll simplify by in this section considering embeddings which depend only on $\rho$, so that $t'=0$ and our entangling region is spacelike. The Lagrangian

    \begin{equation}\label{eqn: Lagrangian r}
        \mathcal{L} = r^2 \rho^2 \sqrt{f(r)r^2 \rho'^2+1}
    \end{equation}

    leads to the following equation of motion for $\rho(r)$:

    \begin{align}\label{eqn: EoM}
        2r^2\rho(r) - 4 r^3 f(r)\rho(r)^2\rho'(r) - r^4 f'(r)\rho(r)^2\rho'(r) +  2r^4 f(r)\rho(r)\rho'(r)^2 \nonumber \\ -  r^4 f(r)\rho(r)^2\rho''(r) - 3r^5 f(r)^2 \rho(r)^2\rho'(r)^3 - \frac{1}{2} r^6 f(r)f'(r)\rho(r)^2 \rho'(r)^3 = 0
    \end{align}

    where primes stand for differentiation with respect to $r$. We switch to dimensionless variables:

    \begin{equation}
        u = \frac{r_*}{r} \quad\quad X = \frac{\mu}{r_*^3} \quad\quad Y = \frac{2\tilde{g}^2}{9} r_*^2 \quad\quad Z = \frac{c^2}{r_*^3}
    \end{equation}

    so the AdS boundary $r \rightarrow \infty$ is now at $u=0$ and the cigar ends at $u=1$. The function $f(r)$ becomes:

    \begin{equation}
        f(u) = -X u^3 + \frac{Y}{u^2}(1- Zu^3)^2.
    \end{equation}

    Now we can rewrite the equation of motion, equation \ref{eqn: EoM}, in the dimensionless variables as:

    \begin{equation}\label{eqn: diff_eqn}
        (f\rho^2)\rho''+(\frac{1}{2}u^2ff'\rho^2 - 3uf^2\rho^2)\rho'^3 - (2  f\rho)\rho'^2 + ( f'\rho^2-\frac{2  }{u}f\rho^2)\rho' - \frac{2}{u^2}\rho = 0
    \end{equation}

    where primes now stand for differentiation with respect to $u$. We want to solve this to find the embedding $\rho(u)$. There will be two different types of solution, depending on whether the surface reaches the end of space $r=r_*$ or not. The geometry of the 8-surface looks like $S_\rho \times S_\phi \times M$ so we have two circles -- one depending on the radius of the embedding surface in the $\rho$ direction, and one depending on the radius of the cigar (in the $\phi$ direction) -- and some other manifold $M$. The two types of solution are distinguished by which circle shrinks and smoothly ends first. \\

    The boundary conditions for small spheres (those with embeddings that end before the confining scale) will be:

    \begin{equation}\label{eqn: small spheres BCs}
        \rho(u \rightarrow 0) = R \quad\quad\quad\quad \Big[ \frac{d\rho}{du}|_{u=u_0} \Big]^{-1} = 0
    \end{equation}

    with the second condition ensuring regularity of the embedding at the turning point.

    We follow the procedure of \cite{Jokela:2025cyz} and use the following series expansion about the turning point as an ansatz to numerically solve equation \ref{eqn: diff_eqn} (for small sphere embeddings):

    \begin{equation}\label{eqn: small ansatz}
        \rho(u) = \sum^\infty_{k=1} b_k (u_0 - u)^{k/2}
    \end{equation}

    with the numerical coefficients $b_k$ determined by requiring that equation \ref{eqn: diff_eqn} is solved at each order in the series.

    For large spheres (those with embeddings ending at the confining scale) the boundary conditions will be:

    \begin{equation}
        \rho(u \rightarrow 0) = R \quad\quad\quad\quad \rho(u=1) = \rho_0
    \end{equation}

    with $\rho_0$ the radius of the embedding in the $\rho$ direction at the confining scale.

\begin{figure}
    	\centering
    	\includegraphics[width = 4in]{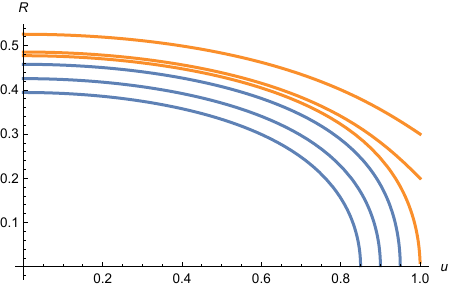}
    	\caption{Small sphere (blue) and large sphere (orange) embeddings, for a range of $u_0$ and $\rho_0$ values. We have made the parameter choice $(X,Y,Z) = (1,4,1/2)$. Here we only plot half of the embeddings, with the other half being the reflection in the horizontal axis. The horizontal axis is the radial coordinate $u$ with the AdS boundary at $u=0$.}
   	\label{fig: embeddings}
    \end{figure}

    We use the following series expansion about $u=1$:

    \begin{equation}\label{eqn: large ansatz}
        \rho(u) = \rho_0 + \sum_{k=1}^{\infty} c_k (1 - u)^k
    \end{equation}

    to numerically solve equation \ref{eqn: diff_eqn}. Examples of the two types of solution for spacelike entangling regions are shown in Figure \ref{fig: embeddings}. \\

    The condition $f(r_*) = 0$ written in terms of $u$ and the dimensionless parameters becomes $-X + Y(1-Z)^2 = 0$ so there is a constraint on the allowed choices for $X$, $Y$, and $Z$. Values not satisfying this equation cannot descend from any arbitrary choice of the original $\tilde{g}$, $\mu$, and $c$ parameters. \\

    \subsection{Entanglement entropy}

    Now that we have solved the equation of motion and found the minimal embedding surfaces, we can compute the entanglement entropy. The EE, written using the dimensionless parameters, is:

    \begin{equation}\label{eqn: unregulated EE}
        4 G_{10}S_{EE} = - \hat{\mathcal{N}} \int^0_{u_0} du \ \frac{r_*^3}{u^4} \rho^2 \sqrt{f(u) u^2 \rho'^2 + 1}
    \end{equation}

    where $u_0 = r_*/r_0$ is the turning point of the surface in the dimensionless coordinate $u$, and the $\rho$ and $\rho'$ come from the embeddings that we've just found, which minimise this quantity. When $u$ is close to 0, $\rho$ becomes constant and $\rho'$ tends to 0, and we can see that the integrand diverges. This necessitates the definition of a regularised entanglement entropy, in analogy with equation \ref{eqn: EE} for slab regions. We use the following series expansion for the embedding to find the behaviour close to $u=0$:

    \begin{equation}\label{eqn: log expansion}
        \rho(u) = \sum_{k=0}^{\infty}\sum_{l=0}^k a_{kl} (\ln u)^l u^k = a_{00} - \frac{1}{3Y a_{00}}u^2 - a_{40} u^4 + \mathcal{O}(u^6)
    \end{equation}

    with the $a_{kl}$ coefficients being constants depending on the parameters used to generate each embedding. We can eliminate most of the coefficients $a_{kl}$ by requiring that diverging terms vanish, and we then substitute the resulting series expansion into the Lagrangian:

    \begin{equation}
        \mathcal{L}_u = \frac{r_*^3}{u^4} \rho^2 \sqrt{f(u) u^2 \rho'^2 + 1}
    \end{equation}

    which is just \ref{eqn: Lagrangian r} after the coordinate switch. We find that close to $u=0$ the Lagrangian is dominated by the following diverging terms:

    \begin{equation}
        \mathcal{L}_u = r_*^3 \left(\frac{a_{00}^2}{u^4} - \frac{4}{9Yu^2} + \mathcal{O}(1) \right)
    \end{equation}

    so for the regularised entanglement entropy we should use:

    \begin{equation}
        \frac{4 G_{10} S_{EE}}{\hat{\mathcal{N}}} = -  \int^0_{u_0}du \ \frac{r_*^3}{u^4}\rho^2 \sqrt{f(u)u^2\rho'^2 + 1} + \int^0_1 du \ r_*^3 \Big( \frac{a_{00}^2}{u^4} - \frac{4}{9Y u^2} \Big).
    \end{equation}

\begin{figure}
    	\centering
    	\includegraphics[width = 6.5in]{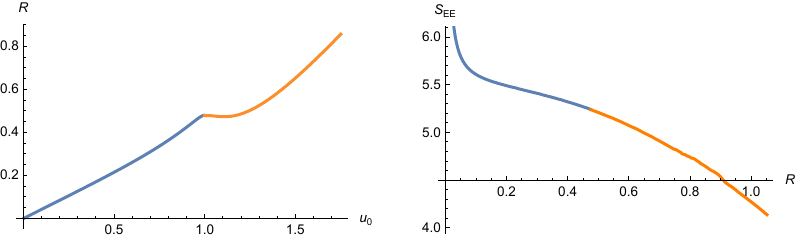}
    	\caption{The embedding radius (the radius of the ball-shaped entangling region on the AdS boundary) $R$ as a function of the turning point $u_0$, and the entanglement entropy as a function of the embedding radius. The blue part comes from small sphere embeddings, and the orange from large spheres. When plotting the orange part we instead vary $\rho_0$ (which corresponds to embeddings with $u_0 > 1$) but plot with with the same scale, so this part of the plot should only be interpreted qualitatively. The embedding radius increases monotonically with $u_0$ for small spheres, but then dips before increasing again for large spheres. Based on this we might expect a swallowtail shape close to the point where the regimes switch, but this is not visible in the plot. Zooming in reveals that the orange line does double back on itself briefly before continuing down, as expected.}
   	\label{fig: spherical EE}
    \end{figure}

    We can find $a_{00}$ (which will be a function of $u_0$ or $\rho_0$) from our numerical solutions, where it is just the radius of the sphere at the boundary $u=0$. The first of the diverging terms can then be seen to originate from embeddings which hang straight down from the boundary to the confining scale. Plotting the entanglement entropy against the embedding radius $R$ (or $a_{00}$) we obtain Figure \ref{fig: spherical EE}.

\subsection{Liu-Mezei c-function}

    The Liu-Mezei (LM) c-function \cite{Liu:2013una,Liu:2012eea} is a proposal for a c-function which can be constructed from the entanglement entropy of ball-shaped entangling regions. In 3 dimensions it has been proven to be monotonic as a function of the entangling region radius \cite{Casini:2012ei}. The proposal involves applying a differential operator to the expression for the EE which cancels out the diverging terms. Without regularisation, our expression for the entanglement entropy diverges like:

    \begin{equation}
        S_{EE} \sim \int du \ \mathcal{L}_u \sim - \frac{a_{00}^2}{3 \epsilon^3} + \frac{4 r_*^2}{9 Y \epsilon } + \mathcal{O}(1)
    \end{equation}

    where we have rewritten $u$ close to 0 as $r_* \epsilon$ with $\epsilon$ a small number. This has the expected divergent structure of a $d=5$ cylinder with radius $R$ and length $L$:

    \begin{equation}\label{eqn: cyl divergent structure}
        S(R) = \frac{L}{R} \Big( p_5 \frac{R^3}{\epsilon^3} + p_3 \frac{R}{\epsilon}  + \mathcal{O}(1) \Big)
    \end{equation}

    where $p_5$ and $p_3$ are dimensionless constants. This means that we should actually use the ILN c-function \cite{Ishihara:2012jg}, which is the appropriate one for cylinder-shaped entangling regions, given by

    \begin{equation}
        \mathcal{C}_{\mathrm{ILN}} = \frac{1}{L} \lim_{\epsilon \rightarrow 0} \mathcal{D}^{(d)}_{\mathrm{ball}} (R \partial_R) (R \ S(R))
    \end{equation}

    where $\mathcal{D}^{(d)}_{\mathrm{ball}}$ is the differential operator defining the LM c-function, and is given by

    \begin{equation}
        \mathcal{D}^{(d)}_\mathrm{ball} (R \partial_R ) = \frac{1}{(d-2)!!} \  \begin{cases} (R \partial_R -1)(R \partial_R -3) \ ... \ (R \partial_R - (d-2)) \quad &\mathrm{for} \ d \ \mathrm{odd} \\  R \partial_R (R \partial_R -2) \ ... \ (R \partial_R - (d-2)) \quad\quad \quad \quad &\mathrm{for} \ d \ \mathrm{even}.
        \end{cases}
    \end{equation}

    The $d=5$ c-function therefore takes the form:

    \begin{equation}\label{eqn: ILN c func}
        \mathcal{C}_{\mathrm{ILN}}(R) = \frac{1}{L} \lim_{\epsilon \rightarrow 0} \mathcal{D}^{(5)}_{\mathrm{ball}} (R \partial_R) (R \ S(R)) = \frac{1}{3L} \Big( R^3 \partial_R^2 S(R) - R^2 \partial_R S(R) \Big)
    \end{equation}

\begin{figure}
    	\centering
    	\includegraphics[width = 4in]{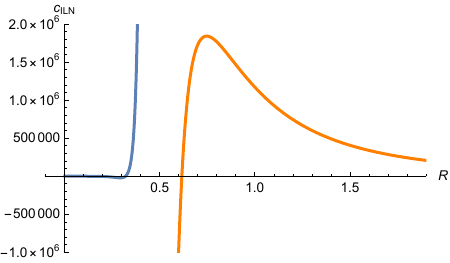}
    	\caption{The ILN c-function from equation \ref{eqn: ILN c func}, applied to the un-regulated part of the entanglement entropy in \ref{eqn: unregulated EE}. Not visible in the plot, the c-function tends to a value in the UV (close to $R=0$) of $\mathcal{C}_{ILN} \approx 0.017$.}
   	\label{fig: ILN c func}
    \end{figure}

    The c-function is constructed in such a way that it removes the UV-divergent terms from the general expression \ref{eqn: cyl divergent structure}. We therefore want to substitute an un-regulated expression for the EE -- this is given by equation \ref{eqn: unregulated EE}. The most important property that we would like for $\mathcal{C}_\mathrm{ILN}(R)$ to obey is monotonicity, but for physical consistency with the dual theory it should also vanish for $u=1$ and tend to a constant value as $u\rightarrow0$. We plot the ILN c-function in Figure \ref{fig: ILN c func}. Immediately we can see that monotonicity is violated, which means that in this situation the proposal \ref{eqn: ILN c func} is not a good c-function. However it does realise the expected asymptotic behaviour -- zooming in close to $R=0$ the function tends to a small positive constant, and extending the plot to larger $R$ allows us to achieve values of $\mathcal{C}_\mathrm{ILN}$ arbitrarily close to zero.

    It was suggested in \cite{Jokela:2025cyz} that the non-monotonicity is linked to the presence of a phase transition. Our above result seems to support this claim. It is interesting that despite this the c-functions in Sections \ref{sec: c function} and \ref{sec: flow central charge} are well-behaved and monotonic.

\subsection{Time-like embeddings}

    For timelike embeddings we repeat the procedure outlined above, but replace the $\rho^2 d \Omega_2^2$ part of the metric in \ref{eqn: ball metric} by $t^2 d \Omega_2^2$. The embedding surfaces again extend in $(\Omega_2, r, \phi, \theta, \varphi, \sigma, \eta)$ and are parametrised by $t=t(r)$ and $\rho = \rho(r)$ but this time we make the choice $\rho'(r) = 0$ so that our entangling region is timelike. We use the same series expansions in equations \ref{eqn: small ansatz} and \ref{eqn: large ansatz} for small and large sphere embeddings (but this time for $t(u)$), and we generate embeddings using the same choice of $(X,Y,Z)=(1,4,1/2)$. \\

\begin{figure}
    	\centering
    	\includegraphics[width = 6in]{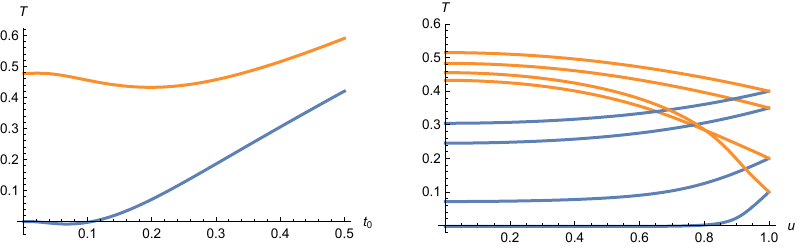}
    	\caption{In the left plot, the embedding radius $T$ as a function of the complex-valued turning point which we vary as $t_0(1+i)$, and in the right plot some examples of timelike embedding surfaces for large spheres. In both plots blue/orange represent the real/imaginary parts of the complex-valued quantities. We can see from the left plot that multiple choices of $t_0$ lead to the same $T$ value. This is reflected in the right plot, where we observe a pair of embeddings which cross in the bulk. As in the spacelike case this means we should expect a phase transition in the EE.}
   	\label{fig: timelike embeddings}
    \end{figure}

    The main difference from the spacelike case is that now our embeddings are in general complex-valued. We consider only embedding profiles along the real $u$ direction, in contrast to \cite{Nunez:2025ppd}. For small spheres we find embeddings identical to the spacelike case, but taking purely imaginary values so that $T$ and $u_0$ are imaginary too. When we go on to choose $t_0$ (the timelike analogue of $\rho_0$) values to generate large sphere embeddings we are free to take $t_0 \in \mathbb{C}$, and we find in general that the embeddings have both real and imaginary parts. When $t_0$ is pure imaginary the embeddings are of identical form to the spacelike case, but are pure imaginary rather then real. When we give $t_0$ both real and imaginary components we find that the imaginary component increases from the end of space while the real component decreases. This is in agreement with the series expansion close to $u=0$ which, in the timelike case, reads:
    
    \begin{equation}\label{eqn: timelike series expansion}
        t(u) = a_{00} + \frac{1}{3Y a_{00}} u^2 + a_{40} u^4 + \mathcal{O}(u^6) 
    \end{equation}
    
    so we see that for $a_{00} \in \mathbb{C}$ the imaginary part of the $u^2$ term decreases from the boundary while the real part increases. We plot some example embeddings in Figure \ref{fig: timelike embeddings}. \\
    
\begin{figure}
    	\centering
    	\includegraphics[width = 4in]{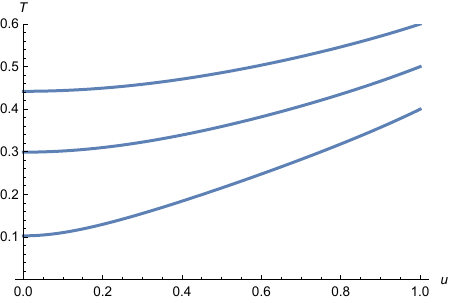}
    	\caption{Real-valued timelike embeddings. As expected from the series expansion \ref{eqn: timelike series expansion} the embeddings increase from the boundary in contrast to the spacelike case. Note that these cannot be obtained by inverting the large sphere embeddings in the spacelike case (shown in Figure \ref{fig: embeddings}). As a result we should expect a qualitatively different EE.}
   	\label{fig: real timelike embeddings}
    \end{figure}

    Based on the results of Section \ref{sec: reality of TP} we might expect real-valued embeddings to be obtainable for large spheres (which have turning points beyond the end of space). And indeed we find such embeddings, shown in Figure \ref{fig: real timelike embeddings}, which seem to appear only after taking $t_0$ sufficiently large -- in our case this critical value happens to be around $t_0 \approx 0.38$. \\

    Calculating the entanglement entropy again requires identifying the divergent terms. This time we use the same series expansion ansatz, equation \ref{eqn: log expansion}, but we find that close to $u=0$ one of the diverging terms now comes with the opposite sign:

    \begin{equation}
        \mathcal{L}_u = r_*^3 \left(\frac{a_{00}^2}{u^4} + \frac{4}{9 Yu^2} + \mathcal{O}(1) \right).
    \end{equation}

    We use this, as before, to define the regularised entanglement entropy:

    \begin{equation}\label{eqn: regularised timelike EE}
        \frac{4 G_{10} S_{EE}}{\hat{\mathcal{N}}} = -  \int^0_{u_0}du \ \frac{r_*^3}{u^4}t^2 \sqrt{1-f(u)u^2t'^2} + \int^0_1 du \ r_*^3 \Big( \frac{a_{00}^2}{u^4} + \frac{4}{9Y u^2} \Big).
    \end{equation}

\begin{figure}
    	\centering
    	\includegraphics[width = 4in]{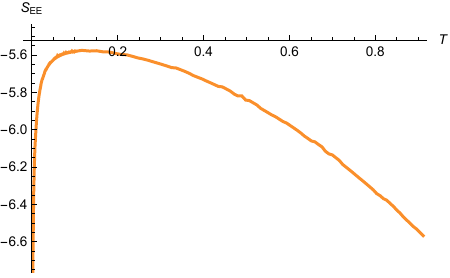}
    	\caption{The real-valued timelike entanglement entropy obtained from the large sphere embeddings shown in Figure \ref{fig: real timelike embeddings}. As in the spacelike case we observe an overall downwards trend for large $T$, but the EE diverges near the critical value $t_0 \approx 0.38$ corresponding to zero embedding radius.}
   	\label{fig: timelike EE}
    \end{figure}

    The calculation involves integrating over our embedding profiles, which in general are complex-valued, as we have seen. This means that the entanglement entropy in general will be complex-valued, and should more properly be interpreted as a pseudo-entropy \cite{Doi:2022iyj,Nakata:2020luh}. One way of achieving a real-valued entanglement entropy is to consider only the imaginary part of our embeddings. For pure imaginary turning points this just switches us back to the spacelike case (with an overall minus sign), as can be seen from equation \ref{eqn: regularised timelike EE} where two factors of $i$ from the $t'^2$ change the sign inside the square root. However, since we have real embeddings for sufficiently large $t_0$, we can use these to obtain real-valued EE in the usual way (without having to pick by hand which part of the embedding to use). We plot this entanglement entropy in Figure \ref{fig: timelike EE}, and find that it diverges close to the critical point corresponding to $t_0 \approx 0.38$.

\section{Conclusion}\label{sec: conclusion}

    In this work we have computed holographic entanglement entropy for spacelike and timelike entangling regions, considering both slab-shaped and ball-shaped entangling regions, for an infinite class of supergravity backgrounds. The Type IIB backgrounds presented provide a smooth geometry with a well-defined brane interpretation dual to a flow between 5-dimensional quiver theory SCFTs and confining 4-dimensional QFTs. Computing the entanglement entropy we found multiple embedding branches, and swallowtail-like structures which signal the presence of phase transitions. We also obtained monotonic c-functions (difficult to obtain from curved entangling regions in dimensions greater than three \cite{Jokela:2025cyz}) which decrease from the UV to the IR, consistent with RG flow expectations, and which become zero at a finite value of $r$, reflecting the gapped nature of the dual theory. Both the c-function derived from the entanglement entropy and the flow central charge display these features, and we therefore find that the transition from 5d to 4d effective theory is reflected in both the geometry and the EE. In particular we should highlight that the c-function from equation \ref{eqn: c function} which we obtain from the EE of slab regions is a covariant c-function for the flow between dimensions -- it would be interesting to compare with other models and check whether this behaviour still holds.\\

    The analysis provides a comprehensive, top-down treatment which connects geometry, brane physics, and quantum information measures, and offers a framework which can also be applied to other backgrounds possessing different RG flows, possibly including theories which are both supersymmetric and confining. It also invites further exploration of higher-dimensional holographic timelike EE, phase transitions, and entanglement c-theorems. \\
    
    Future directions for research could include investigating backgrounds with deformations that preserve some supersymmetry, in contrast to the SUSY-breaking example presented here, and extending the analysis to anisotropic entangling regions which could be a mixture of spacelike and timelike regions, or could probe the internal directions of the spacetime as in \cite{Chatzis:2025wfv}. Another possible avenue might be to consider whether phase transitions in the EE can be related to other observables, eg. Wilson loops or correlation lengths.

\section*{Acknowledgements}
    I would like to thank Carlos N\'u\~nez in particular for suggesting this project to me, and for his continued help, guidance, and support throughout. I would also like to thank Nicolò Bragagnolo, Dimitrios Chatzis, and Madison Hammond for many helpful discussions. This work was supported by the STFC grant no. ST/Y509644/1.


\begin{thebibliography}{99}


\bibitem{Maldacena:1997re}
J.~M.~Maldacena,
``The Large $N$ limit of superconformal field theories and supergravity,''
Adv. Theor. Math. Phys. \textbf{2} (1998), 231-252
[\arXiv{hep-th/9711200} [hep-th]].

\bibitem{Gubser:1998bc}
S.~S.~Gubser, I.~R.~Klebanov and A.~M.~Polyakov,
``Gauge theory correlators from noncritical string theory,''
Phys. Lett. B \textbf{428}, 105-114 (1998)
[\arXiv{hep-th/9802109} [hep-th]].

\bibitem{Witten:1998qj}
E.~Witten,
``Anti de Sitter space and holography,''
Adv. Theor. Math. Phys. \textbf{2}, 253-291 (1998)
[\arXiv{hep-th/9802150} [hep-th]].

\bibitem{Witten:1998zw}
E.~Witten,
``Anti-de Sitter space, thermal phase transition, and confinement in gauge theories,''
Adv. Theor. Math. Phys. \textbf{2} (1998), 505-532
[\arXiv{hep-th/9803131} [hep-th]].

\bibitem{Polchinski:2000uf}
J.~Polchinski and M.~J.~Strassler,
``The String dual of a confining four-dimensional gauge theory,''
[\arXiv{hep-th/0003136} [hep-th]].

\bibitem{Klebanov:2000hb}
I.~R.~Klebanov and M.~J.~Strassler,
``Supergravity and a confining gauge theory: Duality cascades and chi SB resolution of naked singularities,''
JHEP \textbf{08} (2000), 052
[\arXiv{hep-th/0007191} [hep-th]].

\bibitem{Maldacena:2000yy}
J.~M.~Maldacena and C.~Nunez,
``Towards the large N limit of pure N=1 superYang-Mills,''
Phys. Rev. Lett. \textbf{86} (2001), 588-591
[\arXiv{hep-th/0008001} [hep-th]].

\bibitem{Romans:1986er}
L.~J.~Romans,
``Selfduality for Interacting Fields: Covariant Field Equations for Six-dimensional Chiral Supergravities,''
Nucl. Phys. B \textbf{276} (1986), 71

\bibitem{Legramandi:2021uds}
A.~Legramandi and C.~Nunez,
``Electrostatic description of five-dimensional SCFTs,''
Nucl. Phys. B \textbf{974} (2022), 115630
[\arXiv{2104.11240} [hep-th]].

\bibitem{Ryu:2006bv}
S.~Ryu and T.~Takayanagi,
``Holographic derivation of entanglement entropy from AdS/CFT,''
Phys. Rev. Lett. \textbf{96} (2006), 181602
[\arXiv{hep-th/0603001} [hep-th]].

\bibitem{Klebanov:2007ws}
I.~R.~Klebanov, D.~Kutasov and A.~Murugan,
``Entanglement as a probe of confinement,''
Nucl. Phys. B \textbf{796} (2008), 274-293
[\arXiv{0709.2140} [hep-th]].

\bibitem{Kol:2014nqa}
U.~Kol, C.~Nunez, D.~Schofield, J.~Sonnenschein and M.~Warschawski,
``Confinement, Phase Transitions and non-Locality in the Entanglement Entropy,''
JHEP \textbf{06} (2014), 005
[\arXiv{1403.2721} [hep-th]].

\bibitem{Jain:2020rbb}
P.~Jain and S.~Mahapatra,
``Mixed state entanglement measures as probe for confinement,''
Phys. Rev. D \textbf{102} (2020), 126022
[\arXiv{2010.07702} [hep-th]].

\bibitem{Rangamani:2016dms}
M.~Rangamani and T.~Takayanagi,
``Holographic Entanglement Entropy,''
Lect. Notes Phys. \textbf{931} (2017), pp.1-246
Springer, 2017,
[\arXiv{1609.01287} [hep-th]].

\bibitem{Doi:2023zaf}
K.~Doi, J.~Harper, A.~Mollabashi, T.~Takayanagi and Y.~Taki,
``Timelike entanglement entropy,''
JHEP \textbf{05} (2023), 052
[\arXiv{2302.11695} [hep-th]].

\bibitem{Doi:2022iyj}
K.~Doi, J.~Harper, A.~Mollabashi, T.~Takayanagi and Y.~Taki,
``Pseudoentropy in dS/CFT and Timelike Entanglement Entropy,''
Phys. Rev. Lett. \textbf{130} (2023) no.3, 031601
[\arXiv{2210.09457} [hep-th]].

\bibitem{Guo:2025pru}
W.~z.~Guo and J.~Xu,
``Duality of Ryu-Takayanagi surfaces inside and outside the horizon,''
Phys. Rev. D \textbf{112} (2025) no.10, L101901
[\arXiv{2502.16774} [hep-th]].

\bibitem{Afrasiar:2024lsi}
M.~Afrasiar, J.~K.~Basak and D.~Giataganas,
``Timelike entanglement entropy and phase transitions in non-conformal theories,''
JHEP \textbf{07} (2024), 243
[\arXiv{2404.01393} [hep-th]].

\bibitem{Nunez:2025gxq}
C.~Nunez and D.~Roychowdhury,
``Timelike entanglement entropy: A top-down approach,''
Phys. Rev. D \textbf{112} (2025) no.2, 026030
[\arXiv{2505.20388} [hep-th]].

\bibitem{Jena:2024tly}
S.~S.~Jena and S.~Mahapatra,
``A note on the holographic time-like entanglement entropy in Lifshitz theory,''
JHEP \textbf{01} (2025), 055
[\arXiv{2410.00384} [hep-th]].

\bibitem{Nakata:2020luh}
Y.~Nakata, T.~Takayanagi, Y.~Taki, K.~Tamaoka and Z.~Wei,
``New holographic generalization of entanglement entropy,''
Phys. Rev. D \textbf{103} (2021) no.2, 026005
[\arXiv{2005.13801} [hep-th]].

\bibitem{Heller:2024whi}
M.~P.~Heller, F.~Ori and A.~Serantes,
``Geometric Interpretation of Timelike Entanglement Entropy,''
Phys. Rev. Lett. \textbf{134} (2025) no.13, 131601
[\arXiv{2408.15752} [hep-th]].

\bibitem{Heller:2025kvp}
M.~P.~Heller, F.~Ori and A.~Serantes,
``Temporal Entanglement from Holographic Entanglement Entropy,''
Phys. Rev. X \textbf{15} (2025) no.4, 041022
[\arXiv{2507.17847} [hep-th]].

\bibitem{Fatemiabhari:2024aua}
A.~Fatemiabhari and C.~Nunez,
``From conformal to confining field theories using holography,''
JHEP \textbf{03} (2024), 160
[\arXiv{2401.04158} [hep-th]].

\bibitem{Anabalon:2021tua}
A.~Anabalon and S.~F.~Ross,
``Supersymmetric solitons and a degeneracy of solutions in AdS/CFT,''
JHEP \textbf{07} (2021), 015
[\arXiv{2104.14572} [hep-th]].

\bibitem{Anabalon:2024che}
A.~Anabal{\'o}n, H.~Nastase and M.~Oyarzo,
``Supersymmetric AdS solitons and the interconnection of different vacua of $ \mathcal{N} $ = 4 Super Yang-Mills,''
JHEP \textbf{05} (2024), 217
[\arXiv{2402.18482} [hep-th]].

\bibitem{Chatzis:2025dnu}
D.~Chatzis, M.~Hammond, G.~Itsios, C.~Nunez and D.~Zoakos,
``Universal observables, SUSY RG-flows and holography,''
JHEP \textbf{08} (2025), 134
[\arXiv{2506.10062} [hep-th]].

\bibitem{Chatzis:2024kdu}
D.~Chatzis, A.~Fatemiabhari, C.~Nunez and P.~Weck,
``SCFT deformations via uplifted solitons,''
Nucl. Phys. B \textbf{1006} (2024), 116659
[\arXiv{2406.01685} [hep-th]].

\bibitem{Nunez:2023nnl}
C.~Nunez, M.~Oyarzo and R.~Stuardo,
``Confinement in (1 + 1) dimensions: a holographic perspective from I-branes,''
JHEP \textbf{09} (2023), 201
[\arXiv{2307.04783} [hep-th]].

\bibitem{Nunez:2023xgl}
C.~Nunez, M.~Oyarzo and R.~Stuardo,
``Confinement and D5-branes,''
JHEP \textbf{03} (2024), 080
[\arXiv{2311.17998} [hep-th]].

\bibitem{Kumar:2024pcz}
S.~P.~Kumar and R.~Stuardo,
``Twisted circle compactification of $ \mathcal{N} $ = 4 SYM and its holographic dual,''
JHEP \textbf{08} (2024), 089
[\arXiv{2405.03739} [hep-th]].

\bibitem{Barbosa:2024smw}
M.~Barbosa, H.~Nastase, C.~Nunez and R.~Stuardo,
``Penrose limits of I-branes, twist-compactified D5-branes, and spin chains,''
Phys. Rev. D \textbf{110} (2024) no.4, 046015
[\arXiv{2405.08767} [hep-th]].

\bibitem{Macpherson:2024qfi}
N.~T.~Macpherson, P.~Merrikin and R.~Stuardo,
``Circle compactifications of Minkowski$_{D}$ solutions, flux vacua and solitonic branes,''
JHEP \textbf{08} (2025), 143
[\arXiv{2412.15102} [hep-th]].

\bibitem{Macpherson:2025pqi}
N.~T.~Macpherson, P.~Merrikin, C.~Nunez and R.~Stuardo,
``Twisted-circle compactifications of SQCD-like theories and holography,''
JHEP \textbf{08} (2025), 146
[\arXiv{2506.15778} [hep-th]].

\bibitem{Castellani:2024pmx}
F.~Castellani,
``TsT-Generated Solutions in Type IIB Supergravity from Twisted Compactification of AdS$_{5}${\texttimes}T$^{1,1}$,''
JHEP \textbf{04} (2025), 055
[\arXiv{2411.04199} [hep-th]].

\bibitem{Castellani:2024ial}
F.~Castellani and C.~Nunez,
``Holography for confined and deformed theories: TsT-generated solutions in type IIB supergravity,''
JHEP \textbf{12} (2024), 155
[\arXiv{2410.00094} [hep-th]].

\bibitem{Legramandi:2021aqv}
A.~Legramandi and C.~Nunez,
``Holographic description of SCFT$_{5}$ compactifications,''
JHEP \textbf{02} (2022), 010
[\arXiv{2109.11554} [hep-th]].

\bibitem{Hanany:1996ie}
A.~Hanany and E.~Witten,
``Type IIB superstrings, BPS monopoles, and three-dimensional gauge dynamics,''
Nucl. Phys. B \textbf{492} (1997), 152-190
[\arXiv{hep-th/9611230} [hep-th]].

\bibitem{Nunez:2010sf}
C.~Nunez, A.~Paredes and A.~V.~Ramallo,
``Unquenched Flavor in the Gauge/Gravity Correspondence,''
Adv. High Energy Phys. \textbf{2010} (2010), 196714
[\arXiv{1002.1088} [hep-th]].

\bibitem{Intriligator:1997pq}
K.~A.~Intriligator, D.~R.~Morrison and N.~Seiberg,
``Five-dimensional supersymmetric gauge theories and degenerations of Calabi-Yau spaces,''
Nucl. Phys. B \textbf{497} (1997), 56-100
[\arXiv{hep-th/9702198} [hep-th]].

\bibitem{Ryu:2006ef}
S.~Ryu and T.~Takayanagi,
``Aspects of Holographic Entanglement Entropy,''
JHEP \textbf{08} (2006), 045
[\arXiv{hep-th/0605073} [hep-th]].

\bibitem{Nunez:2009da}
C.~Nunez, M.~Piai and A.~Rago,
``Wilson Loops in string duals of Walking and Flavored Systems,''
Phys. Rev. D \textbf{81} (2010), 086001
[\arXiv{0909.0748} [hep-th]].

\bibitem{Jokela:2020wgs}
N.~Jokela and J.~G.~Subils,
``Is entanglement a probe of confinement?,''
JHEP \textbf{02}, 147 (2021)
[\arXiv{2010.09392} [hep-th]].

\bibitem{Nunez:2025ppd}
C.~Nunez and D.~Roychowdhury,
``Interpolating between spacelike and timelike entanglement via holography,''
Phys. Rev. D \textbf{112} (2025) no.8, L081902
[\arXiv{2507.17805} [hep-th]].

\bibitem{Nunez:2025puk}
C.~Nunez and D.~Roychowdhury,
``Holographic timelike entanglement across dimensions,''
JHEP \textbf{11} (2025), 100
[\arXiv{2508.13266} [hep-th]].

\bibitem{Lieb:1973cp}
E.~H.~Lieb and M.~B.~Ruskai,
``Proof of the strong subadditivity of quantum-mechanical entropy,''
J. Math. Phys. \textbf{14} (1973), 1938-1941

\bibitem{Casini:2017vbe}
H.~Casini, E.~Test{\'e} and G.~Torroba,
``Markov Property of the Conformal Field Theory Vacuum and the a Theorem,''
Phys. Rev. Lett. \textbf{118} (2017) no.26, 261602
[\arXiv{1704.01870} [hep-th]].

\bibitem{Myers:2012ed}
R.~C.~Myers and A.~Singh,
``Comments on Holographic Entanglement Entropy and RG Flows,''
JHEP \textbf{04} (2012), 122
[\arXiv{1202.2068} [hep-th]].

\bibitem{Jokela:2025cyz}
N.~Jokela, J.~Kastikainen, C.~Nunez, J.~M.~Pen{\'\i}n, H.~Ruotsalainen and J.~G.~Subils,
``On entanglement c-functions in confining gauge field theories,''
JHEP \textbf{11} (2025), 101
[\arXiv{2505.14397} [hep-th]].

\bibitem{Bea:2015fja}
Y.~Bea, J.~D.~Edelstein, G.~Itsios, K.~S.~Kooner, C.~Nunez, D.~Schofield and J.~A.~Sierra-Garcia,
``Compactifications of the Klebanov-Witten CFT and new AdS$_{3}$ backgrounds,''
JHEP \textbf{05} (2015), 062
[\arXiv{1503.07527} [hep-th]].

\bibitem{Merrikin:2022yho}
P.~Merrikin, C.~Nunez and R.~Stuardo,
``Compactification of 6d N=(1,0) quivers, 4d SCFTs and their holographic dual Massive IIA backgrounds,''
Nucl. Phys. B \textbf{996} (2023), 116356
[\arXiv{2210.02458} [hep-th]].

\bibitem{Liu:2013una}
H.~Liu and M.~Mezei,
``Probing renormalization group flows using entanglement entropy,''
JHEP \textbf{01} (2014), 098
[\arXiv{1309.6935} [hep-th]].

\bibitem{Liu:2012eea}
H.~Liu and M.~Mezei,
``A Refinement of entanglement entropy and the number of degrees of freedom,''
JHEP \textbf{04} (2013), 162
[\arXiv{1202.2070} [hep-th]].

\bibitem{Casini:2012ei}
H.~Casini and M.~Huerta,
``On the RG running of the entanglement entropy of a circle,''
Phys. Rev. D \textbf{85} (2012), 125016
[\arXiv{1202.5650} [hep-th]].

\bibitem{Ishihara:2012jg}
M.~Ishihara, F.~L.~Lin and B.~Ning,
``Refined Holographic Entanglement Entropy for the AdS Solitons and AdS black Holes,''
Nucl. Phys. B \textbf{872}, 392-426 (2013)
[\arXiv:1203.6153 [hep-th]].

\bibitem{Chatzis:2025wfv}
D.~Chatzis, A.~Fatemiabhari, M.~Giliberti and M.~Hammond,
``Holographic Entanglement Entropy in Quiver Theories,''
[\arXiv{2509.19434} [hep-th]].



\end{thebibliography}
\end{document}